\documentclass[final,authoryear,3p,times,twocolumn]{elsarticle}

\usepackage[latin2]{inputenc}
\usepackage{graphicx}

\usepackage{amssymb}
\usepackage{amsthm}
\usepackage{ulem}

\journal{Applied Radiation and Isotopes 118(2016)266}

\begin{document}

\begin{frontmatter}



\title{Cross section measurement of alpha particle induced nuclear reactions on natural cadmium up to 52 MeV}


\author[1]{F. Ditr\'oi\corref{*}}
\author[1]{S. Tak\'acs}
\author[2]{H. Haba}
\author[2]{Y. Komori}
\author[3]{M. Aikawa}

\cortext[*]{Corresponding author: ditroi@atomki.hu}

\address[1]{Institute for Nuclear Research, Hungarian Academy of Sciences (ATOMKI),  Debrecen, Hungary}
\address[2]{Nishina Center for Accelerator-Based Science, RIKEN, Wako, Japan}
\address[3]{Faculty of Science, Hokkaido University, Sapporo, Japan}

\begin{abstract}
\noindent Cross sections of alpha particle induced nuclear reactions have been measured on thin natural cadmium targets foils in the energy range from 11 to 51.2 MeV. This work was a part of our systematic study on excitation functions of light ion induced nuclear reactions on different target materials. Regarding the cross sections, the alpha induced reactions are not deeply enough investigated. Some of the produced isotopes are of medical interest, others have application in research and industry. The radioisotope $^{117m}$Sn is a very important theranostic (therapeutic + diagnostic) radioisotope, so special care was taken to the results for that isotope. The well-established stacked foil technique followed by gamma-spectrometry with HPGe gamma spectrometers were used. The target and monitor foils in the stack were commercial high purity metal foils. From the irradiated targets $^{117m}$Sn, $^{113}$Sn, $^{110}$Sn, $^{117m,g}$In, $^{116m}$In, $^{115m}$In, $^{114m}$In, $^{113m}$In, $^{111}$In, $^{110m,g}$In, $^{109m}$In, $^{108m}$In, $^{115g}$Cd and $^{111m}$Cd were identified and their excitation functions were derived. The results were compared with the data of the previous measurements from the literature and with the results of the theoretical nuclear reaction model code calculations TALYS 1.8 (TENDL-2015) and EMPIRE 3.2 (Malta). From the cross section curves thick target yields were calculated and compared with the available literature data. 
\end{abstract}


\end{frontmatter}


\section{Introduction}
\label{1}
Cadmium is a metal used in industry as an alloying element in different construction materials and as plating element for corrosion resistance. Its natural occurrence consists of 8 stable isotopes with the composition of $^{106}$Cd: 0.0125, $^{108}$Cd: 0.0089, $^{110}$Cd: 0.1251, $^{111}$Cd: 0.1281, $^{112}$Cd: 0.2413, $^{113}$Cd: 0.1222, $^{114}$Cd: 0.2872, and $^{116}$Cd: 0.0747. Its toxicity limits the biological applicability, but some compounds might be important also in biology. It is also used to control neutron flux in nuclear fission reactors \citep{Intercd}. Our interest for cadmium was induced by its use as target material in order to produce some medically and industrially important radioisotopes. Among these radioisotopes the most important is the so called theranostic (therapeutic + diagnostic) radioisotope $^{117m}$Sn \citep{Baum,dit2016, Stevenson}, but it has also non-oncological use in the medicine \citep{Simon,Stevenson}. The diagnostic applicability is based on its good detectable gamma-lines and reasonable half-life (T$_{1/2}$ = 13.76 d) (Knapp et al., 1983; NuDat, 2014). Besides the reactor production \citep{Mausner}, which results in low specific activity final product \citep{Srivastava} $^{117m}$Sn can also be produced by high energy protons on antimony target resulting in NCA (non-carrier-added) end product \citep{Ermolaev}. The most promising route of the production is the high energy alpha particle induced reaction on natural cadmium or rather enriched $^{116}$Cd target material \citep{Stevenson}. 
$^{111g}$In is used for diagnostic medicine for labeling antibodies or blood cell components, as $\gamma$-emitters for SPECT studies \citep{Qaim2012}. This radioisotope is mainly produced on enriched $^{112}$Cd by using proton irradiation in the energy range up to 25 MeV. The use of $^{111}$In for Auger-electron therapy is also considered recently \citep{Qaim2012}. The series of other radioisotopes investigated in our present experiment are usually emerged by medical radioisotope production as by-product or contamination and that is why the knowledge of their excitation function and yield is almost as important as that of the designed radioisotopes. 
From the industrial point of view, as a radioactive tracer, those product radioisotopes could be interesting, which have suitable long half-life, proper gamma-energy with intense radiation and good production parameters (high yield within the energy range of the available accelerator). Taking into account the above conditions, $^{113}$Sn and $^{111}$In can be good candidates. A measurement on a natural target is important, because any enriched target material contains all natural components in a different ratio and radioisotopes produced from them are contaminants in the final products. For estimation of the contaminating amount of different radioisotopes it is necessary to use the experimental results on natural target composition.
Earlier studies of alpha particle induced reactions mainly targeted the production of medically important radioisotopes, such as $^{117m}$Sn \citep{Andreyev,dit2016,Fukushima,Knapp,Qaim1984} but there are also studies giving experimental data for all detectable radio-products \citep{Adam,Hermanne,Qaim2015,Qaim1984}.

\section{Experimental}
\label{2}

Our Cd targets were high-purity commercial (Goodfellow) foils with a measured thickness of 15.6 $\mu$m. The composition of the stack is shown in Fig. 1. The stack was assembled from 11 identical groups of Cd-Cd-Ti target foils, where the 10.9 $\mu$m Ti foils are served both as monitor and as catcher foils for the radioactive recoil products from the preceding Cd foil. The double Cd foil configuration was also used to be able to correct for the recoil products. The irradiation was performed at the beam line of the AVF cyclotron of RIKEN RI Beam Factory by using an E$_{\alpha}$ = 51.2 MeV alpha beam. The beam energy was determined by Time of Flight (TOF) method \citep{Watanabe2015, Watanabe2014}. The irradiation time was slightly more than one hour and the beam current did not exceed 50 nA and it was kept constant within 5\% during the whole irradiation. The collected charge was recorded in every minute in order to determine the total collected charge as well as for confirmation of the beam stability. After a short cooling time the stack was disassembled and the gamma-spectra of the target foils were measured by using ORTEC HPGe gamma spectrometers with the corresponding electronics and evaluation software. The target foils were measured several times to catch both the short-lived (short cooling time and short measuring time) and the long-lived (longer cooling time and longer measuring time) radioactive products. After evaluation also a recoil correction was made by using the data of the products caught in the Ti foils. The evaluated results for the monitor foils are shown in Fig. 2 based on the $^{nat}$Ti($\alpha$,x)$^{51}$Cr monitor reaction by comparing our evaluated cross section values with the upgraded recommended curve from the IAEA monitor reaction database \citep{Tarkanyi}. Figure 2 shows a good agreement between our monitor data and the recommended excitation function. The cross sections of the different nuclear reactions were assessed by using the well-known activation formula by using the peak areas, irradiation, cooling and measuring times and the nuclear data of the reactions and isotopes in Table 1 \citep{Nudat}. 
The uncertainties of the calculated data were determined by taking square root from the sum in quadrature of all individual contributions (International-Bureau-of-Weights-and-Measures, 1993): number of incident particles (5\%), target thickness (2\%), detector efficiency (5\%), nuclear data (3\%), peak areas (1-20\%). The average uncertainty in the assessed cross sections was than 8-20\%.

\begin{figure}
\includegraphics[width=0.5\textwidth]{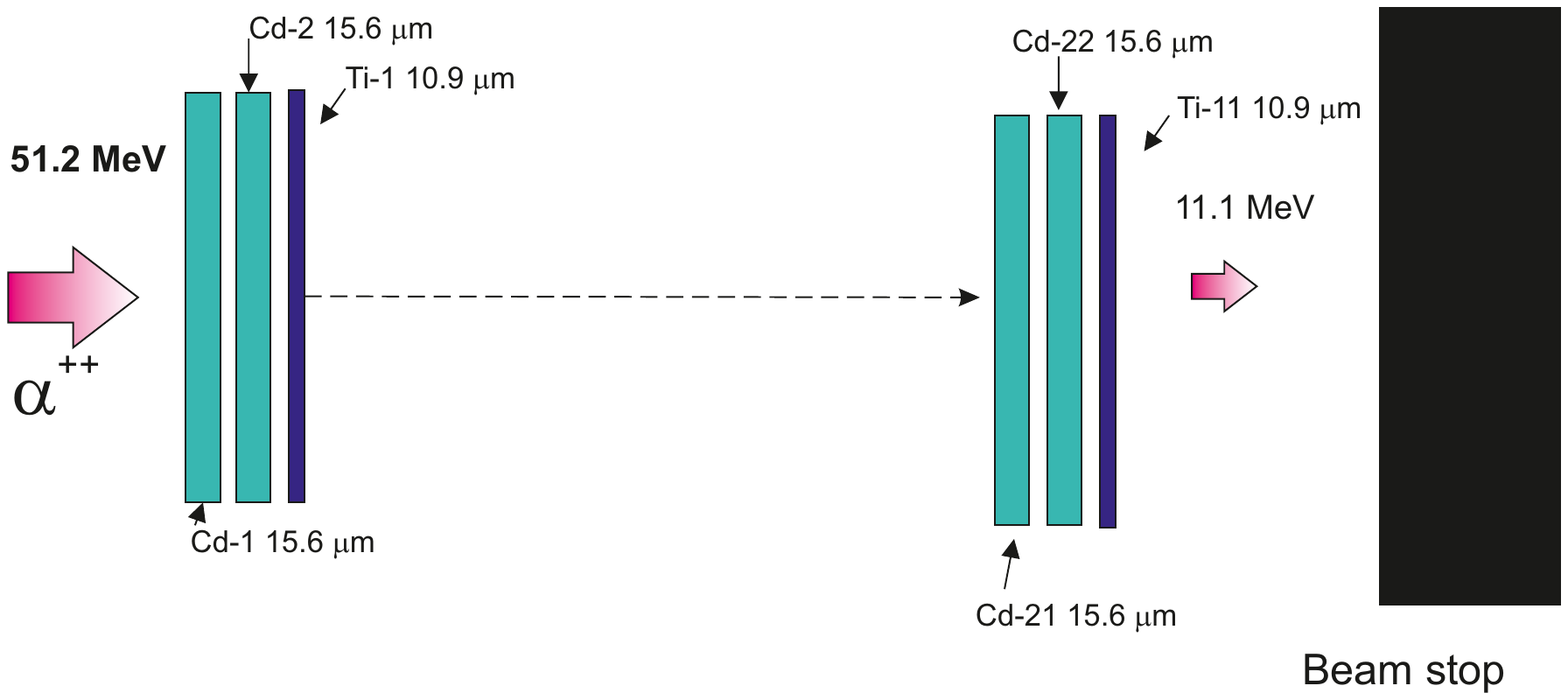}
\caption{Schematic set-up of the irradiated stack}
\label{fig:2}       
\end{figure}

\begin{figure}
\includegraphics[width=0.5\textwidth]{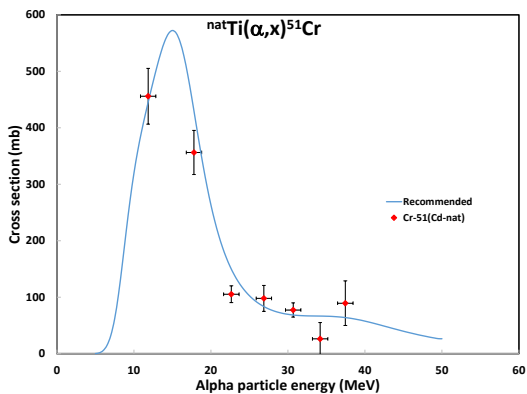}
\caption{Comparison the results on our monitor foils with the recommended cross section of the $^{nat}$Ti($\alpha$,x)$^{51}$Cr reaction	}
\label{fig:2}       
\end{figure}

\begin{table*}[t]
\tiny
\caption{Nuclear data for the radioisotopes in this experiment \citep{Nudat, Pritychenko}}
\begin{center}
\begin{tabular}{|p{0.55in}|p{0.3in}|p{0.4in}|p{0.2in}|p{0.2in}|p{0.6in}|p{0.3in}|} \hline 
\textbf{Isotope\newline spin\newline level energy(keV)} & \textbf{Half-life\newline } & \textbf{Decay mode\newline (\%)\newline } & \textbf{E$_{\gamma}$\newline (keV)\newline } & \textbf{I$_{\gamma}$\newline (\%)\newline } & \textbf{Contributing reactions\newline } & \textbf{Threshold\newline (MeV)\newline } \\ \hline 
$^{117m}$Sn & 14 d & IT 100 & 156.02 & 2.113 & $^{116}$Cd($\alpha$,3n) & 21.11 \\ \hline 
11/2$^{-}$ &  &  & 158.56 & 86.4 & $^{114}$Cd($\alpha$,n) & 5.76 \\ \hline 
314.58 & ~ & ~ & ~ & ~ & $^{117}$In decay & ~ \\ \hline 
$^{113}$Sn & 115.09 d & $\varepsilon$ 100 & 391.70 & 64.97 & $^{110}$Cd($\alpha$,n) & 7.95 \\ \hline 
1/2$^{+}$ &  &  &  &  & $^{111}$Cd($\alpha$,2n) & 15.17 \\ \hline 
 &  &  &  &  & $^{112}$Cd($\alpha$,3n) & 24.90 \\ \hline 
 &  &  &  &  & $^{113}$Cd($\alpha$,4n) & 31.66 \\ \hline 
~ & ~ & ~ & ~ & ~ & $^{114}$Cd($\alpha$,5n) & 41.01 \\ \hline 
$^{110}$Sn & 4.11 h & $\varepsilon$ 100 & 280.46 & 97 & $^{108}$Cd($\alpha$,2n) & 17.76 \\ \hline 
0$^{+}$ &  &  &  &  & $^{110}$Cd($\alpha$,4n) & 35.69 \\ \hline 
 &  &  &  &  & $^{111}$Cd($\alpha$,5n) & 42.83 \\ \hline 
~ &  &  &  &  &  &  \\ \hline 
$^{117m}$In & 116.2 min & IT 47.1 & 315.30 & 19.1 & $^{116}$Cd($\alpha$,2np) & 21.81 \\ \hline 
1/2$^{-}$ &  & $\beta^{-}$ 52.9 &  &  & $^{114}$Cd($\alpha$,p) & 6.46 \\ \hline 
315.30 & ~ & ~ & ~ & ~ & $^{117}$Cd decay & ~ \\ \hline 
${}^{117g}$In & 43.2 min & ${\beta}^{-}$ 100 & 158.6 & 87 & $^{116}$Cd($\alpha$,2np) & 21.49 \\ \hline 
9/2${}^{+}$ &  &  & 552.9 & 100 & ${}^{114}$Cd($\alpha$,p) & 6.14 \\ \hline 
~ & ~ & ~ & ~ & ~ & $^{117}$In decay & ~ \\ \hline 
$^{116m}$In & 54.29 min & ${\beta}^{-}$ 100 & 1097.28 & 58.5 & $^{116}$Cd($\alpha$,3np) & 21.49 \\ \hline 
5$^{+}$ &  &  & 1293.56 & 84.8 & $^{114}$Cd($\alpha$,np) & 15.34 \\ \hline 
127.28 &  &  &  &  & $^{113}$Cd($\alpha$,p) & 5.99 \\ \hline 
 
$^{115m}$In & 4.486 h & IT 95 & 336.24 & 45.8 & ${}^{116}$Cd($\alpha$,4np) & 37.92 \\ \hline 
1/2$^{-}$ &  & ${\beta}^{-}$ 5 &  &  & $^{114}$Cd($\alpha$,2np) & 22.58 \\ \hline 
336.24 &  &  &  &  & $^{113}$Cd($\alpha$,np) & 13.22 \\ \hline 
 &  &  &  &  & ${}^{112}$Cd($\alpha$,p) & 6.45 \\ \hline 
 & ~ & ~ & ~ & ~ & ${}^{115}$Cd decay & ~ \\ \hline 
$^{114m}$In & 49.51 d & IT 100 & 190.27 & 15.56 & ${}^{111}$Cd($\alpha$,p) & 5.94 \\ \hline 
5$^{+}$ &  &  &  &  & $^{112}$Cd($\alpha$,np) & 15.67 \\ \hline 
190.27 &  &  &  &  & ${}^{113}$Cd($\alpha$,2np) & 22.43 \\ \hline 
 &  &  &  &  & $^{114}$Cd($\alpha$,3np) & 31.79 \\ \hline 
 &  &  &  &  & $^{116}$Cd($\alpha$,5np) & 47.12 \\ \hline 
 & ~ & ~ & ~ & ~ & ~ & ~ \\ \hline 
$^{113m}$In & 99.476 min & IT 100 & 391.70 & 64.94 & $^{110}$Cd($\alpha$,p) & 6.45 \\ \hline 
1/2$^{+}$ &  &  &  &  & $^{111}$Cd($\alpha$,np) & 13.68 \\ \hline 
391.69 &  &  &  &  & $^{112}$Cd($\alpha$,2np) & 23.40 \\ \hline 
 &  &  &  &  & $^{113}$Cd($\alpha$,3np) & 30.17 \\ \hline 
 &  &  &  &  & $^{114}$Cd($\alpha$,4np) & 39.52 \\ \hline 
 & ~ & ~ & ~ & ~ & ~ & ~ \\ \hline 
$^{111}$In & 2.8047 d & $\varepsilon$ 100 & 171.28 & 90.7 & $^{108}$Cd($\alpha$,p) & 5.94 \\ \hline 
9/2$^{+}$ &  &  & 245.35 & 94.1 & $^{110}$Cd($\alpha$,2np) & 23.80 \\ \hline 
 &  &  &  &  & $^{111}$Cd($\alpha$,3np) & 31.02 \\ \hline 
 &  &  &  &  & $^{112}$Cd($\alpha$,4np) & 40.74 \\ \hline 
~ & ~ & ~ & ~ & ~ & $^{113}$Cd($\alpha$,5np) & 47.50 \\ \hline 
$^{110g}$In & 4.92 h & $\varepsilon$ 100 & 657.75 & 98 & ${}^{108}$Cd($\alpha$,np) & 16.30 \\ \hline 
7$^{+}$ &  &  & 884.68 & 93 & ${}^{110}$Cd(?,3np) & 34.16 \\ \hline 
 &  &  &  &  & $^{111}$Cd($\alpha$,4np) & 41.37 \\ \hline 
~ & ~ & ~ & ~ & ~ & $^{112}$Cd($\alpha$,5np) & 51.09 \\ \hline 
${}^{110m}$In & 69.1 min & $\varepsilon$ 100 & 657.75 & 97.74 & $^{108}$Cd($\alpha$,np) & 16.36 \\ \hline 
2${}^{+}$ &  &  &  &  & $^{110}$Cd($\alpha$,3np) & 34.22 \\ \hline 
62.08 &  &  &  &  & $^{111}$Cd($\alpha$,4np) & 41.43 \\ \hline 
 & ~ & ~ & ~ & ~ & $^{112}$Cd($\alpha$,5np) & 51.15 \\ \hline 
${}^{109g}$In & 4.167 h & $\varepsilon$ 100 & 203.5 & 73.5 & $^{106}$Cd($\alpha$,p) & 10.53 \\ \hline 
9/2$^{+}$ &  &  &  &  & $^{108}$Cd($\alpha$,2np) & 24.65 \\ \hline 
 &  &  &  &  & $^{110}$Cd($\alpha$,4np) & 42.50 \\ \hline 
~ & ~ & ~ & ~ & ~ & $^{111}$Cd($\alpha$,5np) & 49.72 \\ \hline 
$^{108m}$In & 39.6 min & $\varepsilon$ 100 & 632.9 & 76.4 & $^{106}$Cd($\alpha$,np) & 16.58 \\ \hline 
2$^{+}$ &  &  &  &  & $^{108}$Cd($\alpha$,3np) & 35.51 \\ \hline 
29.76 & ~ & ~ & ~ & ~ & ~ & ~ \\ \hline 
$^{115g}$Cd & 53.46 h & ${\beta}^{-}$ 100 & 527.90 & 27.5 & $^{116}$Cd($\alpha$,3n2p) & 38.45 \\ \hline 
1/2$^{+}$ &  &  &  &  & $^{116}$Cd($\alpha$,n$\alpha$) & 9.18 \\ \hline 
 &  &  &  &  & $^{114}$Cd($\alpha$,n2p) & 22.93 \\ \hline 
 &  &  &  &  & $^{113}$Cd($\alpha$,2p) & 13.58 \\ \hline 
~ & ~ & ~ & ~ & ~ & ${}^{115}$Ag decay & ~ \\ \hline 
$^{115m}$Cd & 44.56 d & ${\beta}^{-}$ 100 & 933.84 & 2 & $^{116}$Cd($\alpha$,3n2p) & 38.63 \\ \hline 
11/2${}^{-}$ &  &  &  &  & $^{116}$Cd($\alpha$,n$\alpha$) & 9.36 \\ \hline 
181.05 &  &  &  &  & $^{114}$Cd($\alpha$,n2p) & 23.11 \\ \hline 
 &  &  &  &  & ${}^{113}$Cd($\alpha$,2p) & 13.76 \\ \hline 
~ & ~ & ~ & ~ & ~ & $^{115}$Ag decay & ~ \\ \hline 
$^{111m}$Cd & 48.5 min & IT 100 & 150.83 & 19.1 & $^{110}$Cd($\alpha$,n2p) & 22.49 \\ \hline 
11/2${}^{-}$ &  &  &  &  & ${}^{111}$Cd($\alpha$,2n2p) & 29.71 \\ \hline 
396.22 &  &  &  &  & ${}^{112}$Cd($\alpha$,3n2p) & 39.43 \\ \hline 
 &  &  &  &  & $^{112}$Cd($\alpha$,n$\alpha$) & 10.13 \\ \hline 
 &  &  &  &  & $^{113}$Cd($\alpha$,4n2p) & 46.19 \\ \hline 
 &  &  &  &  & $^{113}$Cd($\alpha$,2n$\alpha$) & 16.89 \\ \hline 
 &  &  &  &  & $^{114}$Cd($\alpha$,3n$\alpha$) & 26.25 \\ \hline 
 & ~ & ~ & ~ & ~ & $^{116}$Cd($\alpha$,5n$\alpha$) & 41.69 \\ \hline 
\end{tabular}

\end{center}
\end{table*}




\section{Nuclear reaction model code calculations}
\label{3}

Our new experimental data were also compared with the calculated results of two nuclear reaction model codes. TALYS \citep{Koning2007}, recent version 1.8 is available on-line \citep{Koning2015} and its latest calculated results are tabulated in the in the TENDL-2015 on-line library \citep{Koning2013, Koning2015b}. The on-line tabulated version contains adjusted results of the TALYS code. Calculations were also performed with the newest version of the EMPIRE code \citep{Herman2007}, the EMPIRE 3.2 (Malta) \citep{Herman2012}, which uses the latest reference input parameters library RIPL-3 \citep{Capote}. The codes were launched with the default input parameters and all possible reaction channels were considered, including emission of complex particles at our bombarding energies. The level density fits were adjusted after the first run of the EMPIRE 3.2 code. Comparison of the codes output with our new experimental results could help the developers to improve the models behind these calculations.

\section{Excitation functions}
\label{4}
The experimental excitation functions obtained in this experiment are presented graphically in Figs. 3-19 and numerically in Tables 2 and 3. The results are compared with the previous measurements found in the literature and also with the results of the theoretical nuclear model code calculations.

\section{Comparison with nuclear model calculations}
\label{4}
The experimental excitation functions obtained in this experiment are presented graphically in Figs. 3-19 and numerically in Tables 2 and 3. The results are compared with the previous measurements found in the literature and also with the results of the theoretical nuclear model code calculations.

\subsection{The $^{nat}$Cd($\alpha$,x)$^{117m}$Sn reactions}
\label{4.1}
The Sn radioisotopes are produced directly in ($\alpha$,xn) reactions by emission of neutrons. The $^{117m}$Sn radioisotope can only be produced directly from the two highest mass number cadmium isotopes through the $^{114}$Cd($\alpha$,3n) and $^{116}$Cd($\alpha$,n) reactions. It is also formed from the decay of the A = 117 mass number isobar chain. In order to let the A = 117 parents decay, our results were determined from a late measurement series, in such a way our results are cumulative for the $^{117m}$Sn radioisotope (Fig. 3). Both TENDL-2015 and EMPIRE 3.2 follow the trend of our new experimental data. Below 25 MeV the TENDL-2015 approximation is better but above 40 MeV EMPIRE 3.2 gives closer approach. Between 25 and 40 MeV both theoretical codes underestimate slightly our experimental results. The new results are in good agreement with the recent data of \citep{Duchemin} above 32 MeV and slightly higher below 32 MeV. The data of \citep{Khandaker}  are slightly higher than ours below 35 MeV and show acceptable agreement above this energy, except one salient value at 39.5 MeV. The previous experimental works of \citep{Qaim1984} and \citep{Hermanne} report completely different results far away both from our data and from each other. They do not show the trend reproduced by the theoretical model codes.

\begin{figure}
\includegraphics[width=0.5\textwidth]{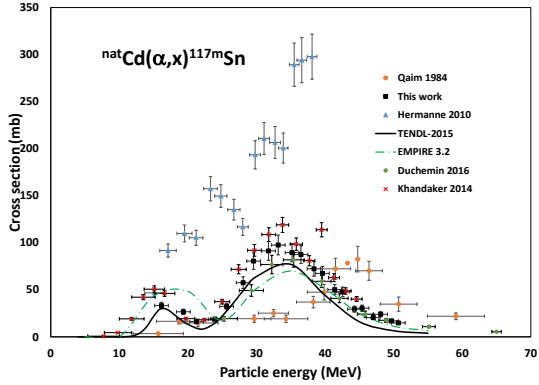}
\caption{Excitation function of the $^{nat}$Cd($\alpha$,x)$^{117m}$Sn nuclear reactions in comparison with the previous experiments from the literature and with the results of the nuclear reaction model code calculations}
\label{fig:3}       
\end{figure}

\subsection{The $^{nat}$Cd($\alpha$,x)$^{113}$Sn reactions}
\label{4.2}
The $^{113}$Sn has two isomeric states, both can be produced directly from the A=110-114 stable cadmium isotopes by ($\alpha$,xn) reactions. Its T$_{1/2}$  =  21.4 min excited state decays partly to the longer-lived ground state, so the spectra measured after the complete decay of the isomeric state were used in our calculations. Our new results are shown in Fig. 4 in comparison with the previous experimental results from the literature and with the theoretical nuclear reaction model code calculations. The new data are in good agreement with the TENDL-2015 prediction taking into account the cumulative cross section. TENDL-2015 could even reproduce the local minimum in the 15-25 MeV energy region. EMPIRE 3.2 gives much lower results and very different shape. The previous experimental data agree neither with the new data nor with each other or with the theoretical predictions. The recent data set of \citep{Duchemin} supports our data above  45 MeV and below 32 MeV, but it cannot reproduce the maximum of the excitation function. The data of \citep{Khandaker} are also lower than our values except for one data point at 16 MeV.

\begin{figure}
\includegraphics[width=0.5\textwidth]{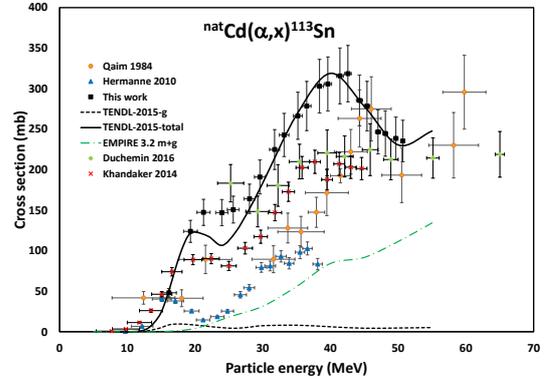}
\caption{Excitation function of the $^{nat}$Cd($\alpha$,x)$^{113}$Sn nuclear reactions in comparison with the previous experiments from the literature and with the results of the nuclear reaction model code calculations}
\label{fig:4}       
\end{figure}

\subsection{The $^{nat}$Cd($\alpha$,x)$^{110}$Sn reactions}
\label{4.3}
The radioisotope $^{110}$Sn has a relatively short half-life (T$_{1/2}$ = 4.11 h), which requires quick measurements with short cooling times. Only direct production is possible from the stable cadmium isotopes in the A = 108-111 mass region in the used energy range. Our results are in excellent agreement with the previous data of \citep{Hermanne} and \citep{Khandaker} (Fig. 5). The data of  (Duchemin et al., 2016) are slightly lower, especially above 40 MeV. The TENDL-2015 prediction is lower below 40 MeV but above that energy becomes higher than our values. EMPIRE 3.2 gives almost the same estimation as TENDL-2015 up to 40 MeV, but strongly declines above this energy. 

\begin{figure}
\includegraphics[width=0.5\textwidth]{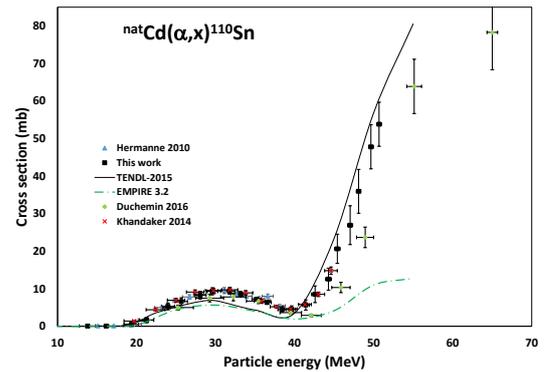}
\caption{Excitation function of the $^{nat}$Cd($\alpha$,x)$^{110}$Sn nuclear reactions in comparison with the previous experiments from the literature and with the results of the nuclear reaction model code calculations}
\label{fig:5}       
\end{figure}

\subsection{The $^{nat}$Cd($\alpha$,x)$^{117m}$In reactions}
\label{4.4}
The direct production of indium isotopes in alpha bombardment of cadmium targets requires also the emission of a proton in an ($\alpha$,pxn) reaction. $^{117m}$In can be produced on $^{114}$Cd and $^{116}$Cd only. The activity of $^{117m}$In can be independently measured through its 315.3 keV gamma-line. This isomeric level can also be fed from the parent $^{117}$Cd radioisotope, which was not detected in our spectra. The new data are presented in Fig. 6 together with the previous experimental results and with the results of TALYS-2015 and EMPIRE 3.2 model code calculations. Because of the low counting statistics both our and the previous data of \citep{Hermanne} are slightly scattered, but show an acceptable agreement and also agree with the TENDL-2015 prediction, while EMPIRE 3.2 gives large overestimation. It is difficult to judge the agreement with the recent data of \citep{Khandaker}, because they report extraordinary large uncertainties.

\begin{figure}
\includegraphics[width=0.5\textwidth]{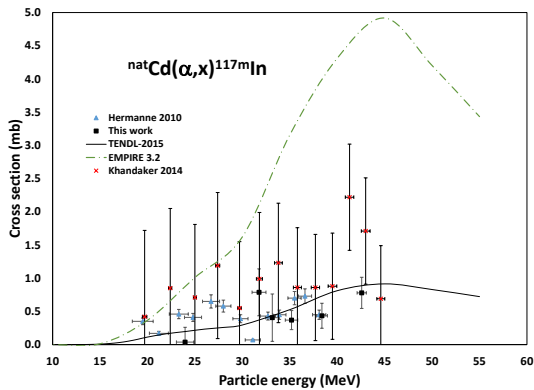}
\caption{Excitation function of the $^{nat}$Cd($\alpha$,x)$^{117m}$In nuclear reactions in comparison with the previous experiments from the literature and with the results of the nuclear reaction model code calculations}
\label{fig:6}       
\end{figure}

\subsection{The $^{nat}$Cd($\alpha$,x)$^{117g}$In reactions}
\label{4.5}
The ground state of $^{117}$In is produced in the same reactions as its excited state (see above), and the ground state is also fed by decay of its excited state. A short cooling time was necessary to decrease the decay corrections from the higher isomeric state. Our new data are shown in Fig. 7 together with the previous experimental results and the results of the model code calculations. Our new data are in good agreement with the previous results of \citep{Hermanne} and with the results of \citep{Khandaker}. EMPIRE 3.2 predicts well the experimental values up to 30 MeV. The TENDL-2015 approximation underestimates the experimental values even if cumulative results are considered.

\begin{figure}
\includegraphics[width=0.5\textwidth]{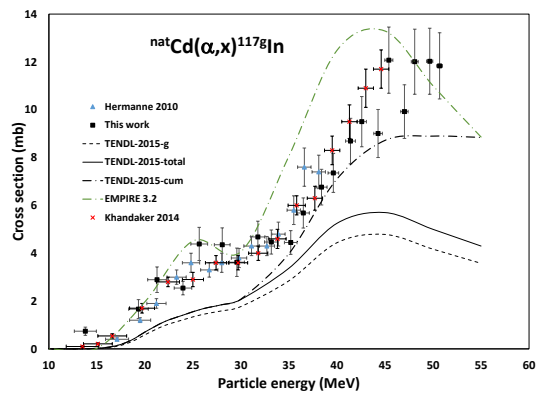}
\caption{Excitation function of the $^{nat}$Cd($\alpha$,x)$^{117g}$In nuclear reactions in comparison with the previous experiments from the literature and with the results of the nuclear reaction model code calculations}
\label{fig:7}       
\end{figure}

\subsection{The $^{nat}$Cd($\alpha$,x)$^{116m2}$In reactions}
\label{4.6}
The radioisotope $^{116}$In can only be produced by direct reactions. The $^{116}$In has a very short-lived (T$_{1/2}$ = 2.18 s) isomeric state, a longer-lived (T$_{1/2}$ = 54.29 min) isomeric state and a very short-lived (T$_{1/2}$ = 14.1 s) ground state. The ground state has a small EC branch to the stable $^{116}$Cd, and the 2.18 s m1 isomeric state decays to the 54.29 min m2 isomeric state, thus the result will be cumulative (m1+m2). Because of the short half-lives only the 54.29 min $^{116m2}$In was detectable for us from the first series of the measurements with short cooling times. The results are presented in Fig. 8 together with the results of the previous experiments from the literature and the results of the theoretical model code calculations. The new data are in excellent agreement with the previous results of \citep{Hermanne, Khandaker}. The EMPIRE 3.2 overestimates while TENDL-2015 underestimates the experimental data significantly.

\begin{figure}
\includegraphics[width=0.5\textwidth]{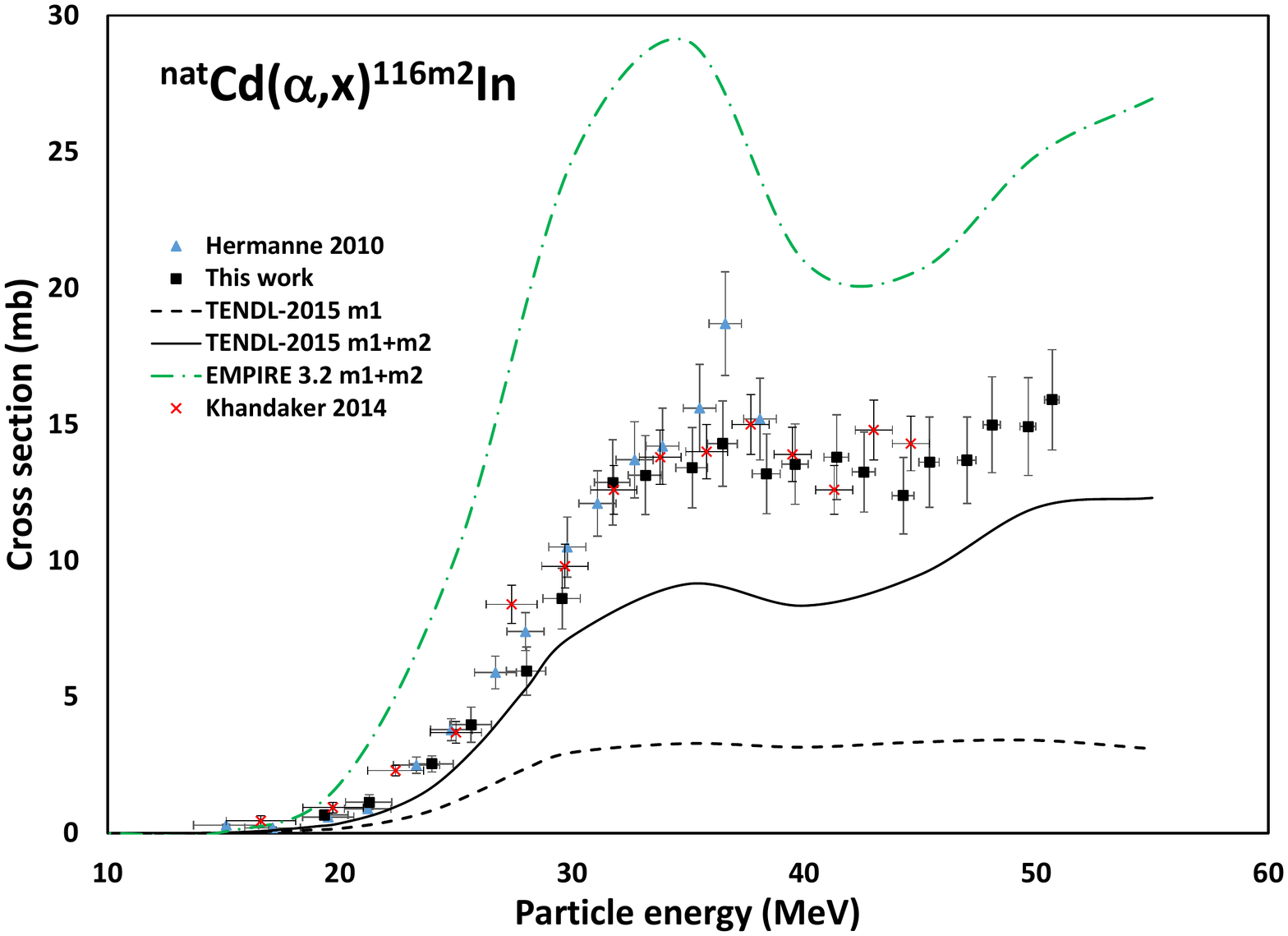}
\caption{Excitation function of the $^{nat}$Cd($\alpha$,x)$^{116m}$In nuclear reactions in comparison with the previous experiments from the literature and with the results of the nuclear reaction model code calculations}
\label{fig:8}       
\end{figure}

\subsection{The $^{nat}$Cd($\alpha$,x)$^{115m}$In reactions}
\label{4.7}
The radioisotope $^{115}$In is produced directly on the stable cadmium isotopes with A = 112-116 in the applied energy range. Decay of the $^{115}$Cd parent isotope also contributes to its production. The $^{115}$Cd decays by 97\% to the almost stable ground state (T$_{1/2}$ = 4.14 1014 a). Because the half-life of the parent isotope is an order of magnitude longer than the daughter of interest, a quick measurement by short cooling time makes the mothers contribution negligible. The results of the present experiment are shown in Fig. 9 together with the previous experimental values and the results of the model code calculations. Our new results are in excellent agreement with the previous data of \citep{Hermanne, Khandaker}. The agreement with the TENDL-2015 curve is also acceptable, but the EMPIRE 3.2 gives much higher values.

\begin{figure}
\includegraphics[width=0.5\textwidth]{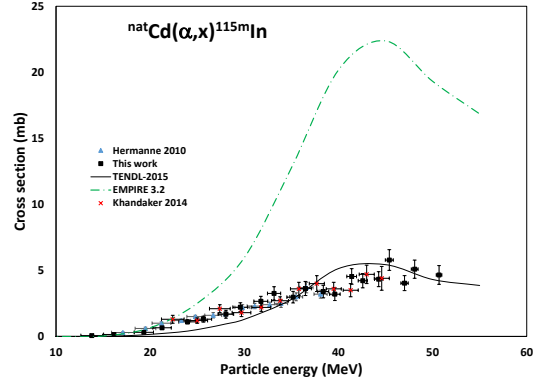}
\caption{Excitation function of the $^{nat}$Cd($\alpha$,x)$^{115m}$In nuclear reactions in comparison with the previous experiments from the literature and with the results of the nuclear reaction model code calculations}
\label{fig:9}       
\end{figure}

\subsection{The $^{nat}$Cd($\alpha$,x)$^{114m}$In reactions}
\label{4.8}
The $^{114}$In can be produced from stable cadmium isotopes with A = 111-116 in direct nuclear reactions with the used bombarding beam energy. There is no parent isotope to feed its population. The relatively long half-life of $^{114m}$In (T$_{1/2}$ = 49.51 d) makes a convenient measurement possible. The results of the present experiment are shown in Fig. 10 together with the previous experimental values and the results of the model code calculations. Our results are in good agreement with the previous experimental data of \citep{Hermanne}. The recent results of \citep{Duchemin} agree with us up to 40 MeV and become smaller at the higher energies. Both our and their dataset are a bit scattered above 40 MeV. The recent data of \citep{Khandaker} are somewhat higher than all the others below 32 MeV. In general, the TENDL-2015 calculation underestimates, while the curve of the EMPIRE 3.2 calculation overestimates the experimental data.

\begin{figure}
\includegraphics[width=0.5\textwidth]{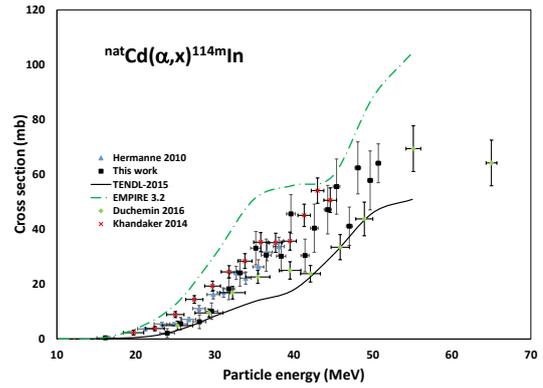}
\caption{Excitation function of the $^{nat}$Cd($\alpha$,x)$^{114m}$In nuclear reactions in comparison with the previous experiments from the literature and with the results of the nuclear reaction model code calculations}
\label{fig:10}       
\end{figure}

\subsection{The $^{nat}$Cd($\alpha$,x)$^{113m}$In reactions}
\label{4.9}
The $^{113m}$In radioisotope can be produced directly from A = 110-114 stable cadmium isotopes in our energy range. The parent isotope $^{113g}$Sn feeds its production, but it was easy to calculate the necessary corrections because their half-lives are very different and the cross section of the $^{113}$Sn is also measured. Our new results are shown in Fig. 11 together with the previous experimental values and the results of the model code calculations. The new data are in excellent agreement with the previous values of \citep{Hermanne, Khandaker} in the whole overlapping region. The TENDL-2015 calculation underestimates the experimental results below 28 MeV, and overestimates above 32 MeV. The EMPIRE 3.2 calculation provides much higher estimation.

\begin{figure}
\includegraphics[width=0.5\textwidth]{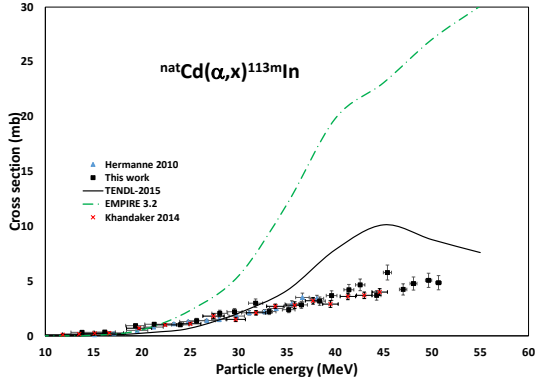}
\caption{Excitation function of the $^{nat}$Cd($\alpha$,x)$^{113m}$In nuclear reactions in comparison with the previous experiments from the literature and with the results of the nuclear reaction model code calculations}
\label{fig:11}       
\end{figure}

\subsection{The $^{nat}$Cd($\alpha$,x)$^{111}$In reactions}
\label{4.10}
The radioisotope $^{111}$In is produced directly from the A = 108-113 stable cadmium isotopes in our energy range. The $^{111m}$Cd decays by 100\% IT to the ground state therefore does not contribute to the production of $^{111}$In. The shorter-lived (T$_{1/2}$ = 35.3 min) $^{111}$Sn decays with $\varepsilon$ = 100\% to $^{111}$In. Spectra measured after the complete decay of the parent radioisotope were used to assess the cumulative cross section. The results are shown in Fig. 12 together with the previous experimental values and the results of the model code calculations. Our new results are in excellent agreement with the previous data of \citep{Hermanne} and \citep{Khandaker}, and are somewhat higher than the recent data of \citep{Duchemin}.  The TENDL-2015 cumulative predictions are in good agreement with the experimental data up to 35 MeV, but they are lower above 36 MeV. Above 36 MeV, the EMPIRE 3.2 declines considerably.

\begin{figure}
\includegraphics[width=0.5\textwidth]{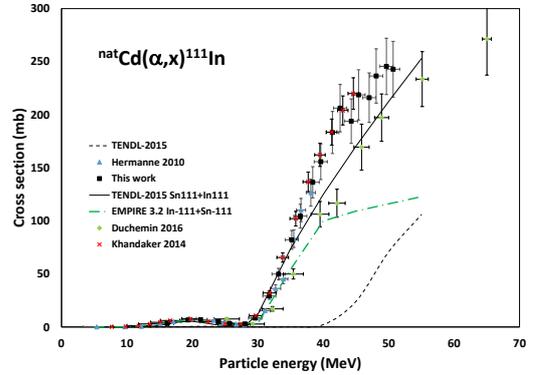}
\caption{Excitation function of the $^{nat}$Cd($\alpha$,x)$^{111}$In nuclear reactions in comparison with the previous experiments from the literature and with the results of the nuclear reaction model code calculations}
\label{fig:12}       
\end{figure}

\subsection{The $^{nat}$Cd($\alpha$,x)$^{110m}$In reactions}
\label{4.11}
The $^{110m}$In is produced by direct nuclear reactions from the A = 108-112 stable cadmium isotopes in the used energy range. Its production is also fed partly by decay of $^{110}$Sn. $^{110}$Sn has longer half-life than $^{110m}$In, so its contribution is unavoidable. The cross section of the $^{110}$Sn production could be measured independently, and in such a way its contribution could be subtracted from the measured activities. The corrected results are shown in Fig. 13 together with the previous experimental values and the results of the model code calculations. Our new results are slightly lower than the previous results of \citep{Hermanne} and in good agreement with the TENDL-2015 predictions. EMPIRE 3.2 also gives acceptable estimation. The data of \citep{Khandaker} gives significantly larger values, especially in the energy range below 40 MeV.

\begin{figure}
\includegraphics[width=0.5\textwidth]{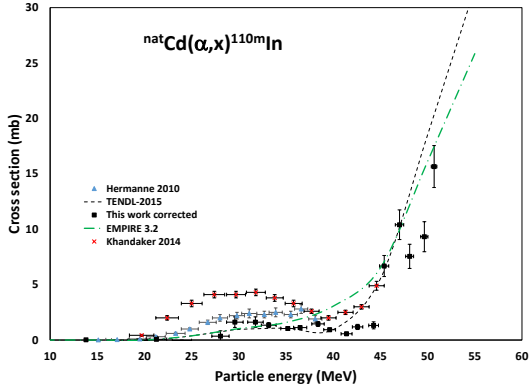}
\caption{Excitation function of the $^{nat}$Cd($\alpha$,x)$^{110m}$In nuclear reactions in comparison with the previous experiments from the literature and with the results of the nuclear reaction model code calculations}
\label{fig:13}       
\end{figure}

\subsection{The $^{nat}$Cd($\alpha$,x)$^{110g}$In reactions}
\label{4.12}
The radioisotope $^{110g}$In can be produced directly on $^{108-112}$Cd at the used bombarding energy. The $^{110m}$In isomeric level decays directly to $^{110}$Cd, and there is no contribution from the $^{110}$Sn decay to $^{110g}$In radioisotope. The results are independent and shown in Fig. 14 together with the previous experimental values and the results of the model code calculations. Our new results are in excellent agreement with the previous data of \citep{Hermanne},  \citep{Khandaker} and \citep{Duchemin}. Both theoretical nuclear reaction model codes agree well up to 42 MeV. Above this energy the TENDL-2015 gives higher, while EMPIRE 3.2 gives lower values.

\begin{figure}
\includegraphics[width=0.5\textwidth]{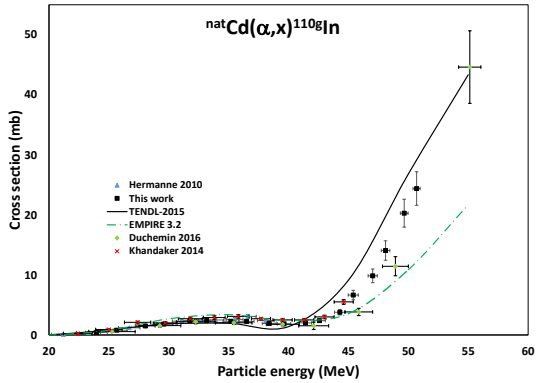}
\caption{Excitation function of the $^{nat}$Cd($\alpha$,x)$^{110g}$In nuclear reactions in comparison with the previous experiments from the literature and with the results of the nuclear reaction model code calculations}
\label{fig:14}       
\end{figure}

\subsection{The $^{nat}$Cd($\alpha$,x)$^{109g}$In reactions}
\label{4.13}
The radioisotope $^{109g}$In can be produced directly on the A = 106-111 stable cadmium isotopes in the used energy range. It has two short-lived metastable states decaying completely onto the ground state. Decay of the parent $^{109}$Sn contributes to its production. The $^{109}$Sn parent isotope is also shorter lived than $^{109g}$In in question. Cumulative production cross sections were determined from the second series of spectra taken after proper cooling time. Our new results are shown in Fig. 15 together with the previous experimental values and the results of the model code calculations. The new results are in good agreement with the previous data of \citep{Hermanne}, except our results are much lower around the local maximum of 20 MeV. The recent data of \citep{Khandaker} are in excellent agreement with our new results and also support our data around the lower energy maximum but seem to have some energy shift at the lower energy points. The recent results of \citep{Duchemin} agree with our data around the 30 MeV local minimum only and could not reproduce the upper maximum at 43 MeV. TENDL-2015 gives relatively good approximation above 32 MeV, while the agreement with the EMPIRE 3.2 calculation result is good up to 32 MeV. 

\begin{figure}
\includegraphics[width=0.5\textwidth]{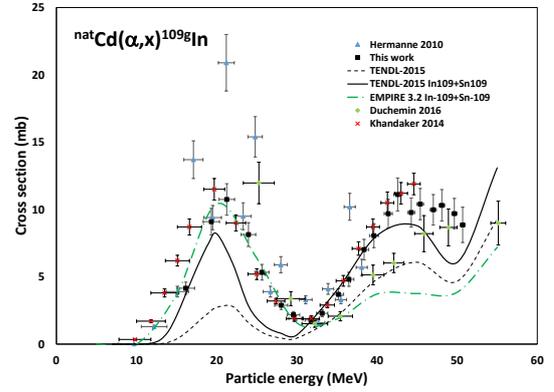}
\caption{Excitation function of the $^{nat}$Cd($\alpha$,x)$^{109g}$In nuclear reactions in comparison with the previous experiments from the literature and with the results of the nuclear reaction model code calculations}
\label{fig:15}       
\end{figure}

\subsection{The $^{nat}$Cd($\alpha$,x)$^{108m}$In reactions}
\label{4.14}
The $^{108m}$In radioisotope is produced directly from the A = 106-108 stable cadmium isotopes in the used energy range. The production of this metastable state of $^{108}$In is also fed from decay of the $^{108}$Sn parent isotope. Because of the short half-life of the parent (T$_{1/2}$ = 10.3 min) and the daughter (T$_{1/2}$ = 39.6 min) it is difficult to separate them based on their half-life difference. Because no gamma-lines of $^{108}$Sn were measured in the present experiment, a correction based on model calculations was possible, which might have caused large uncertainties. The new results are shown in Fig. 16 together with the previous experimental values and the results of the model code calculations. The agreement with the previous data of \citep{Hermanne} is only good around the maximum region, below these energies our data are considerably higher. The recent data of \citep{Khandaker} are lower than our data and those of \citep{Hermanne} above 30 MeV. An agreement with our data is only seen below 25 MeV. The results of the theoretical nuclear model calculations reproduce the trend of the experimental values but give much lower cross sections. 

\begin{figure}
\includegraphics[width=0.5\textwidth]{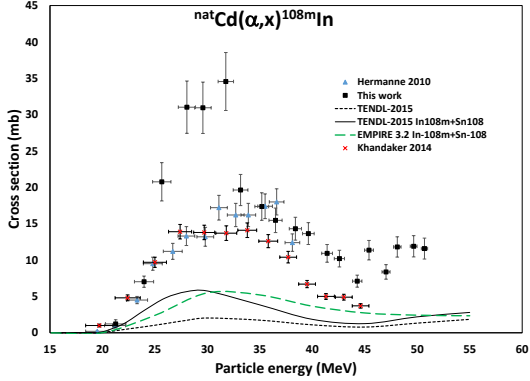}
\caption{Excitation function of the $^{nat}$Cd($\alpha$,x)$^{108m}$In nuclear reactions in comparison with the previous experiments from the literature and with the results of the nuclear reaction model code calculations}
\label{fig:16}       
\end{figure}

\subsection{The $^{nat}$Cd($\alpha$,x)$^{108g}$In reactions}
\label{4.15}
The meta-stable state $^{108m}$In decays only to the levels of $^{108}$Cd, so $^{108g}$In is produced only by direct reactions and by decay of $^{108}$Sn. The half-life of $^{108g}$In is also similar to those of $^{108}$Sn and $^{108m}$In, so for the evaluation of both isomers the same series of measurements can be used. The calculated cross sections are shown in Fig. 17 together with the previous experimental values and the results of the model code calculations. The new results are in good agreement with the previous experimental values of \citep{Hermanne, Khandaker}. The EMPIRE 3.2 model code provides similar shape with somewhat higher maximum value. The TENDL-2015 gives similar trend as the experimental data but higher values in the whole energy region.

\begin{figure}
\includegraphics[width=0.5\textwidth]{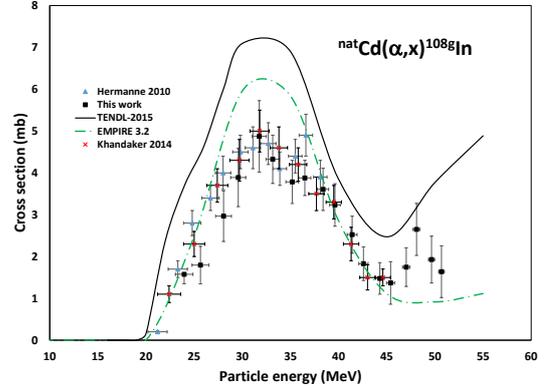}
\caption{Excitation function of the $^{nat}$Cd($\alpha$,x)$^{108g}$In nuclear reactions in comparison with the previous experiments from the literature and with the results of the nuclear reaction model code calculations}
\label{fig:17}       
\end{figure}

\subsection{The $^{nat}$Cd($\alpha$,x)$^{115g}$Cd reactions}
\label{4.16}
The radioisotope $^{115}$Cd can be produced directly from the A = 113-116 stable cadmium isotopes. Gamma photons from decay of the shorter-lived ground-state were detected in our spectra. This isotope can also be produced by decay of the short-lived $^{115}$Ag. Cumulative cross sections were measured after the complete decay of the $^{115}$Ag parent. The new experimental cross sections are shown in Fig. 18 together with the previous experimental values and the results of the model code calculations. Our new results are in acceptable agreement with the previous data of \citep{Hermanne}. The recent results of \citep{Duchemin} and \citep{Khandaker} are somewhat lower and definitely higher above 30 MeV, respectively. Both TENDL-2015 and EMPIRE 3.2 agree with the experimental results only up to 32 MeV and give much lower values above this energy.

\begin{figure}
\includegraphics[width=0.5\textwidth]{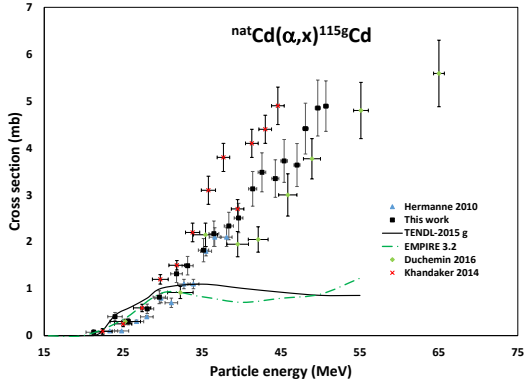}
\caption{Excitation function of the $^{nat}$Cd($\alpha$,x)$^{115g}$Cd nuclear reactions in comparison with the previous experiments from the literature and with the results of the nuclear reaction model code calculations}
\label{fig:18}       
\end{figure}

\subsection{The $^{nat}$Cd($\alpha$,x)$^{111m}$Cd reactions}
\label{4.17}
The radioisotope $^{111m}$Cd can be directly produced from the A = 111-116 stable cadmium isotopes. The feeding mother isotope is $^{111}$Ag, which was not detected in our spectra. Contribution from decay of $^{111}$In to $^{111m}$Cd is negligible. Our new experimental cross sections are shown in Fig. 19 together with the previous experimental values and the results of the model code calculations. Our new results are in good agreement with the previous data of \citep{Hermanne, Khandaker}, except for a couple of salient points. The estimations given by both TENDL-2015 and EMPIRE 3.2 are far away from the experimental results both in shape and in value for most of the investigated energy range. 

\begin{figure}
\includegraphics[width=0.5\textwidth]{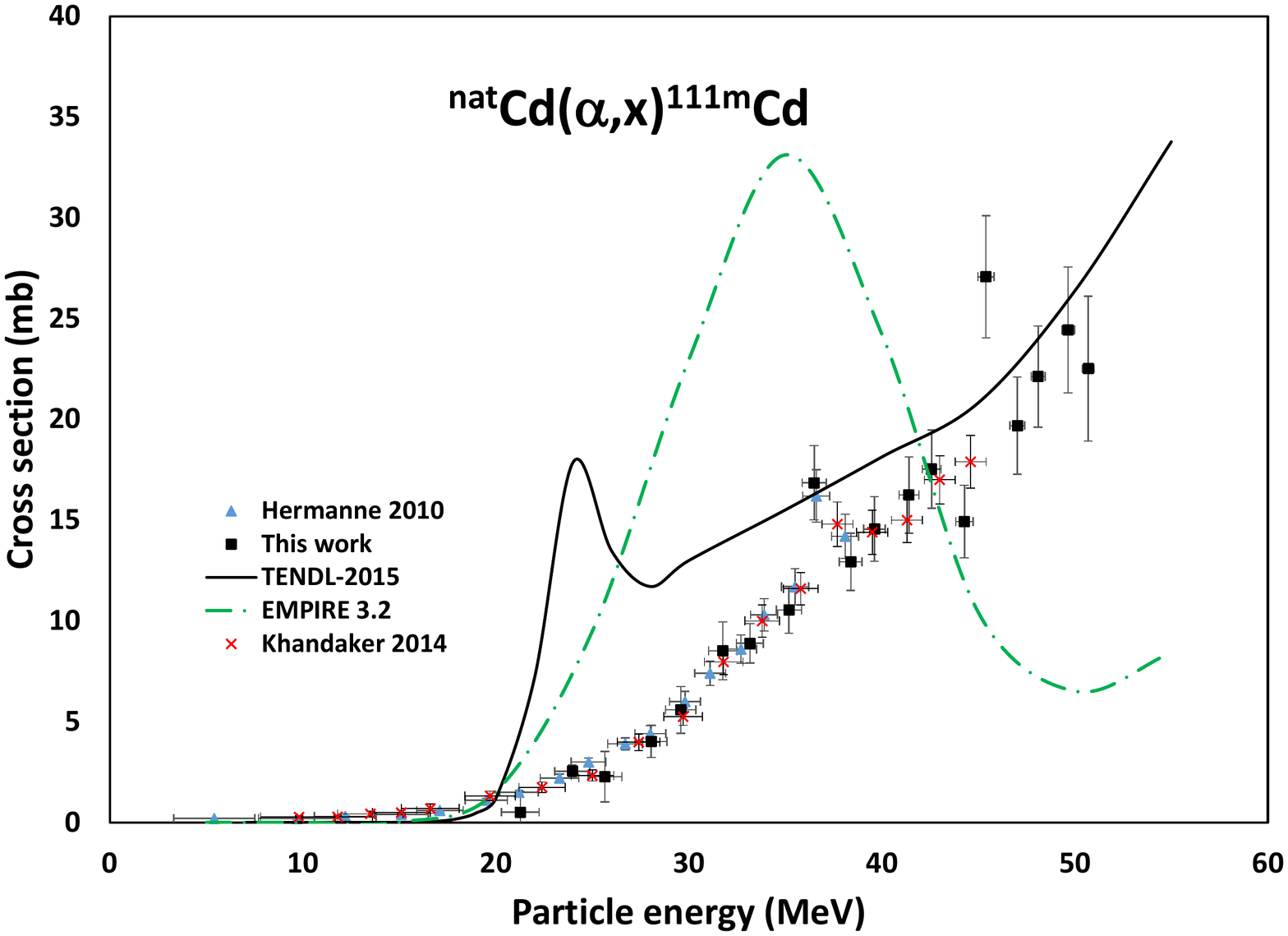}
\caption{Excitation function of the $^{nat}$Cd($\alpha$,x)$^{111}$Cd nuclear reactions in comparison with the previous experiments from the literature and with the results of the nuclear reaction model code calculations}
\label{fig:19}       
\end{figure}

\begin{table*}[t]
\tiny
\caption{Measured cross sections of the $^{nat}$Cd($\alpha$,x)$^{117m,113,110}$Sn, $^{117m,117g,116m,115m,114m}$In nuclear reactions}
\begin{center}
\begin{tabular}{|p{0.15in}|p{0.15in}|p{0.2in}|p{0.2in}|p{0.2in}|p{0.2in}|p{0.2in}|p{0.2in}|p{0.2in}|p{0.2in}|p{0.2in}|p{0.2in}|p{0.2in}|p{0.2in}|p{0.2in}|p{0.2in}|p{0.2in}|p{0.2in}|} \hline 
\multicolumn{2}{|c|}{\textbf{Alpha Energy}} & \multicolumn{2}{|c|}{\textbf{$^{117}$${}^{m}$Sn}} & \multicolumn{2}{|c|}{\textbf{$^{113}$Sn}} & \multicolumn{2}{|c|}{\textbf{$^{110}$Sn}} & \multicolumn{2}{|c|}{\textbf{$^{117m}$In}} & \multicolumn{2}{|c|}{\textbf{$^{117g}$In}} & \multicolumn{2}{|c|}{\textbf{$^{116m}$In}} & \multicolumn{2}{|c|}{\textbf{$^{115m}$In}} & \multicolumn{2}{|c|}{\textbf{$^{114m}$In}} \\ \hline 
\textbf{E} & \textbf{$\Delta$E} & \textbf{${\sigma}$} & \textbf{${\Delta\sigma}$} & \textbf{${\sigma}$} & \textbf{${\Delta\sigma}$} & \textbf{${\sigma}$} & \textbf{${\Delta\sigma}$} & \textbf{${\sigma}$} & \textbf{${\Delta\sigma}$} & \textbf{${\sigma}$} & \textbf{${\Delta\sigma}$} & \textbf{${\sigma}$} & \textbf{${\Delta\sigma}$} & \textbf{${\sigma}$} & \textbf{${\Delta\sigma}$} & \textbf{${\sigma}$} & \textbf{${\Delta\sigma}$} \\ \hline 
\multicolumn{2}{|c|}{\textbf{MeV}} & \multicolumn{16}{|c|}{\textbf{mb}} \\ \hline 
50.7 & 0.3 & 14.9 & 1.6 & 235.4 & 25.5 & 53.8 & 5.8 &  &  & 11.8 & 1.4 & 15.9 & 1.8 & 4.7 & 0.7 & 64.1 & 7.1 \\ \hline 
49.7 & 0.3 & 16.9 & 1.9 & 238.6 & 26.3 & 47.8 & 5.2 & ~ &  & 12.0 & 1.4 & 14.9 & 1.8 & ~ &  & 57.8 & 10.8 \\ \hline 
48.1 & 0.4 & 24.0 & 2.6 & 244.6 & 28.1 & 36.0 & 3.9 & ~ &  & 12.0 & 1.4 & 15.0 & 1.8 & 5.1 & 0.7 & 62.4 & 9.5 \\ \hline 
47.0 & 0.4 & 20.5 & 2.3 & 246.6 & 27.2 & 26.9 & 2.9 & ~ &  & 9.9 & 1.1 & 13.7 & 1.6 & 4.0 & 0.6 & 41.1 & 7.0 \\ \hline 
45.4 & 0.4 & 30.5 & 3.3 & 278.5 & 30.6 & 20.6 & 2.2 & ~ &  & 12.1 & 1.4 & 13.6 & 1.7 & 5.8 & 0.8 & 55.6 & 10.1 \\ \hline 
44.3 & 0.4 & 29.5 & 3.2 & 285.5 & 31.3 & 12.6 & 1.4 & ~ &  & 9.0 & 1.0 & 12.4 & 1.4 & 4.3 & 0.6 & 47.2 & 8.9 \\ \hline 
42.6 & 0.5 & 47.8 & 5.2 & 318.5 & 34.8 & 8.5 & 0.9 & 0.78 & 0.23 & 9.5 & 1.1 & 13.3 & 1.5 & 4.2 & 0.5 & 40.4 & 8.8 \\ \hline 
41.4 & 0.5 & 50.2 & 5.4 & 315.7 & 34.5 & 5.7 & 0.6 & ~ &  & 8.7 & 1.0 & 13.8 & 1.5 & 4.5 & 0.6 & 30.4 & 6.0 \\ \hline 
39.6 & 0.6 & 67.5 & 7.3 & 305.8 & 33.4 & 4.5 & 0.5 & ~ &  & 7.4 & 0.8 & 13.5 & 1.5 & 3.2 & 0.5 & 45.6 & 7.1 \\ \hline 
38.4 & 0.6 & 72.4 & 7.8 & 303.0 & 33.1 & 4.4 & 0.5 & 0.43 & 0.19 & 6.8 & 0.7 & 13.2 & 1.5 & 3.4 & 0.5 & 30.2 & 5.6 \\ \hline 
36.5 & 0.6 & 87.4 & 9.5 & 278.6 & 30.4 & 6.4 & 0.7 & ~ &  & 5.7 & 0.6 & 14.3 & 1.6 & 3.6 & 0.5 & 30.6 & 5.9 \\ \hline 
35.2 & 0.7 & 89.5 & 9.7 & 266.5 & 29.1 & 7.2 & 0.8 & 0.37 & 0.14 & 4.4 & 0.5 & 13.4 & 1.5 & 3.0 & 0.4 & 33.1 & 6.1 \\ \hline 
33.2 & 0.7 & 97.7 & 10.6 & 242.8 & 26.6 & 8.6 & 0.9 & 0.41 & 0.36 & 4.5 & 0.5 & 13.1 & 1.4 & 3.3 & 0.5 & 24.1 & 4.9 \\ \hline 
31.8 & 0.7 & 91.3 & 9.9 & 224.9 & 24.6 & 9.3 & 1.0 & 0.79 & 0.35 & 4.7 & 0.7 & 12.9 & 1.6 & 2.7 & 0.4 & 18.3 & 4.0 \\ \hline 
29.6 & 0.8 & 80.6 & 8.7 & 191.0 & 21.0 & 9.3 & 1.0 & ~ &  & 3.6 & 0.6 & 8.6 & 1.1 & 2.2 & 0.3 & 10.1 & 3.0 \\ \hline 
28.0 & 0.8 & 57.6 & 6.2 & 164.1 & 18.1 & 8.3 & 0.9 & ~ &  & 4.4 & 0.7 & 6.0 & 0.9 & 1.7 & 0.3 & 6.3 & 2.5 \\ \hline 
25.7 & 0.9 & 32.3 & 3.5 & 150.7 & 16.6 & 6.6 & 0.7 & ~ &  & 4.4 & 0.7 & 4.0 & 0.6 & 1.3 & 0.2 & 5.6 & 2.3 \\ \hline 
24.0 & 0.9 & 18.8 & 2.0 & 147.0 & 16.2 & 5.3 & 0.6 & 0.04 & 0.22 & 2.5 & 0.3 & 2.5 & 0.3 & 1.1 & 0.1 & 2.1 & 1.6 \\ \hline 
21.3 & 1.0 & 16.1 & 1.7 & 147.3 & 16.1 & 1.7 & 0.2 & ~ &  & 2.9 & 0.5 & 1.1 & 0.3 & 0.7 & 0.1 & ~ &  \\ \hline 
19.3 & 1.0 & 26.6 & 2.9 & 123.9 & 13.6 & 0.443 & 0.064 & ~ &  & 1.7 & 0.4 & 0.7 & 0.2 & 0.3 & 0.1 & ~ &  \\ \hline 
16.1 & 1.1 & 33.2 & 3.6 & 48.0 & 5.2 & 0.036 & 0.015 & ~ &  & ~ &  & ~ &  & 0.1 & 0.0 & 0.4 & 0.2 \\ \hline 
13.8 & 1.1 &  &  & ~ &  & 0.037 & 0.005 & ~ &  & 0.74 & 0.17 & ~ &  & 0.07 & 0.01 & ~ &  \\ \hline 
\end{tabular}

\end{center}
\end{table*}

\begin{table*}[t]
\tiny
\caption{Measured cross sections of the $^{nat}$Cd($\alpha$,x)$^{113m,110m,110g,109g,108m,108g}$In, $^{115m,111}$Cd nuclear reactions}
\begin{center}
\begin{tabular}{|p{0.15in}|p{0.15in}|p{0.2in}|p{0.2in}|p{0.2in}|p{0.2in}|p{0.2in}|p{0.2in}|p{0.2in}|p{0.2in}|p{0.2in}|p{0.2in}|p{0.2in}|p{0.2in}|p{0.2in}|p{0.2in}|p{0.2in}|p{0.2in}|p{0.2in}|p{0.2in}|} \hline 
\multicolumn{2}{|c|}{\textbf{Alpha Energy}} & \multicolumn{2}{|c|}{\textbf{$^{113m}$In}} & \multicolumn{2}{|c|}{\textbf{$^{111g}$In}} & \multicolumn{2}{|c|}{\textbf{$^{110m}$In}} & \multicolumn{2}{|c|}{\textbf{$^{110g}$In}} & \multicolumn{2}{|c|}{\textbf{$^{109g}$In}} & \multicolumn{2}{|c|}{\textbf{$^{108m}$In}} & \multicolumn{2}{|c|}{\textbf{$^{108g}$In}} & \multicolumn{2}{|c|}{\textbf{$^{115g}$Cd}} & \multicolumn{2}{|c|}{\textbf{$^{111}$${}^{m}$Cd}} \\ \hline 
\textbf{E} & \textbf{$\Delta$E} & \textbf{${\sigma}$} & \textbf{${\Delta\sigma}$} & \textbf{${\sigma}$} & \textbf{${\Delta\sigma}$} & \textbf{${\sigma}$} & \textbf{${\Delta\sigma}$} & \textbf{${\sigma}$} & \textbf{${\Delta\sigma}$} & \textbf{${\sigma}$} & \textbf{${\Delta\sigma}$} & \textbf{${\sigma}$} & \textbf{${\Delta\sigma}$} & \textbf{${\sigma}$} & \textbf{${\Delta\sigma}$} & \textbf{${\sigma}$} & \textbf{${\Delta\sigma}$} & \textbf{${\sigma}$} & \textbf{${\Delta\sigma}$} \\ \hline 
\multicolumn{2}{|c|}{\textbf{MeV}} & \multicolumn{18}{|c|}{\textbf{mb}} \\ \hline 
50.7 & 0.3 & 4.8 & 0.6 & 242.9 & 26.3 & 9.3 & 1.4 & 24.4 & 2.8 & 8.9 & 1.3 & 11.6 & 1.4 & 1.6 & 0.6 & 4.9 & 0.5 & 22.5 & 3.6 \\ \hline 
49.7 & 0.3 & 5.1 & 0.6 & 245.6 & 26.6 & 7.6 & 1.1 & 20.3 & 2.3 & 9.7 & 1.1 & 11.9 & 1.5 & 1.9 & 0.6 & 4.9 & 0.6 & 24.4 & 3.1 \\ \hline 
48.1 & 0.4 & 4.8 & 0.6 & 236.6 & 25.6 & 10.4 & 1.3 & 14.1 & 1.6 & 10.3 & 1.2 & 11.8 & 1.4 & 2.6 & 0.6 & 4.4 & 0.5 & 22.1 & 2.5 \\ \hline 
47.0 & 0.4 & 4.2 & 0.5 & 216.1 & 23.4 & 6.7 & 1.0 & 9.9 & 1.1 & 10.0 & 1.1 & 8.4 & 1.0 & 1.7 & 0.5 & 3.6 & 0.5 & 19.7 & 2.4 \\ \hline 
45.4 & 0.4 & 5.8 & 0.7 & 218.7 & 23.7 & 1.3 & 0.3 & 6.7 & 0.8 & 10.4 & 1.2 & 11.4 & 1.3 & 1.4 & 0.5 & 3.7 & 0.5 & 27.1 & 3.0 \\ \hline 
44.3 & 0.4 & 3.7 & 0.4 & 194.0 & 21.0 & 1.2 & 0.2 & 3.8 & 0.5 & 9.8 & 1.1 & 7.1 & 0.8 & 1.5 & 0.4 & 3.3 & 0.4 & 14.9 & 1.8 \\ \hline 
42.6 & 0.5 & 4.7 & 0.5 & 206.2 & 22.3 & 0.6 & 0.2 & 2.4 & 0.3 & 11.1 & 1.2 & 10.2 & 1.1 & 1.8 & 0.4 & 3.5 & 0.4 & 17.5 & 1.9 \\ \hline 
41.4 & 0.5 & 4.2 & 0.5 & 183.3 & 19.8 & 0.9 & 0.2 & 2.0 & 0.3 & 9.7 & 1.1 & 10.9 & 1.2 & 2.5 & 0.4 & 3.1 & 0.4 & 16.2 & 1.9 \\ \hline 
39.6 & 0.6 & 3.7 & 0.4 & 155.8 & 16.9 & 1.5 & 0.2 & 1.9 & 0.2 & 8.1 & 0.9 & 13.6 & 1.5 & 3.2 & 0.5 & 2.5 & 0.3 & 14.6 & 1.6 \\ \hline 
38.4 & 0.6 & 3.2 & 0.4 & 136.4 & 14.8 & 1.1 & 0.2 & 1.9 & 0.2 & 7.0 & 0.8 & 14.3 & 1.6 & 3.6 & 0.5 & 2.3 & 0.3 & 12.9 & 1.4 \\ \hline 
36.5 & 0.6 & 2.8 & 0.3 & 104.5 & 11.3 & 1.0 & 0.2 & 2.3 & 0.3 & 4.8 & 0.5 & 15.5 & 1.7 & 3.9 & 0.4 & 2.2 & 0.3 & 16.9 & 1.8 \\ \hline 
35.2 & 0.7 & 2.4 & 0.3 & 82.1 & 8.9 & 1.3 & 0.2 & 2.3 & 0.3 & 3.7 & 0.4 & 17.4 & 1.9 & 3.8 & 0.5 & 1.8 & 0.2 & 10.5 & 1.2 \\ \hline 
33.2 & 0.7 & 2.2 & 0.3 & 49.9 & 5.4 & 1.6 & 0.5 & 2.4 & 0.3 & 2.3 & 0.3 & 19.6 & 2.1 & 4.3 & 0.6 & 1.5 & 0.2 & 8.9 & 1.0 \\ \hline 
31.8 & 0.7 & 3.0 & 0.4 & 29.2 & 3.2 & 1.6 & 0.5 & 2.4 & 0.3 & 1.8 & 0.2 & 34.6 & 4.0 & 4.9 & 0.9 & 1.3 & 0.2 & 8.5 & 1.4 \\ \hline 
29.6 & 0.8 & 2.2 & 0.3 & 8.4 & 0.9 & 0.35 & 0.48 & 1.8 & 0.2 & 2.2 & 0.2 & 30.9 & 3.5 & 3.9 & 0.7 & 0.82 & 0.14 & 5.59 & 1.16 \\ \hline 
28.0 & 0.8 & 2.1 & 0.3 & 3.0 & 0.3 & ~ &  & 1.5 & 0.2 & 2.9 & 0.3 & 31.0 & 3.6 & 3.0 & 0.6 & 0.57 & 0.10 & 4.02 & 0.79 \\ \hline 
25.7 & 0.9 & 1.4 & 0.2 & 3.3 & 0.4 & ~ &  & 0.8 & 0.1 & 5.3 & 0.6 & 20.8 & 2.6 & 1.8 & 0.4 & 0.30 & 0.07 & 2.27 & 1.25 \\ \hline 
24.0 & 0.9 & 1.0 & 0.1 & 5.0 & 0.5 & 0.07 & 0.24 & 0.4 & 0.1 & 8.1 & 0.9 & 7.0 & 0.8 & 1.6 & 0.2 & 0.40 & 0.09 & 2.54 & 0.32 \\ \hline 
21.3 & 1.0 & 1.1 & 0.2 & 7.0 & 0.8 & ~ &  & ~ &  & 10.7 & 1.2 & 1.2 & 0.6 & ~ &  & 0.07 & 0.05 & 0.51 & 1.01 \\ \hline 
19.3 & 1.0 & 0.9 & 0.1 & 6.6 & 0.7 & ~ &  & ~ &  & 9.1 & 1.0 & ~ &  & ~ &  & ~ &  &  &  \\ \hline 
16.1 & 1.1 & 0.3 & 0.1 & 3.2 & 0.3 & 0.02 & 0.03 & ~ &  & 4.2 & 0.5 & ~ &  & ~ &  & 0.09 & 0.01 &  &  \\ \hline 
13.8 & 1.1 & 0.32 & 0.04 & ~ &  & ~ &  & ~ &  & ~ &  & ~ &  & ~ &  & ~ &  &  &  \\ \hline 
\end{tabular}

\end{center}
\end{table*}

\section{Thick target yields}
\label{5}

The most important radioisotope from this experiment is $^{117m}$Sn and its physical thick target yield is presented in Fig. 20. Our new data are somewhat higher than the previous results of \citep{Fukushima} and lower than the recent data of (Khandaker et al., 2014). There is another experimental result for the thick target yield in the literature from \citep{Maslov}. They used a limited target thickness at the 35 MeV bombarding energy. We have calculated a yield from our new cross section data for the same target thickness and composition and got a 0.0115 GBq/C thick target yield, which is in acceptable agreement with the data published by Maslov et al.: 0.01042 GBq/C. The same figure presents our new yield data for $^{113}$Sn together with the previous results of \citep{Dmitriev}, which is somewhat lower than our data. For $^{110}$Sn the only previous data is from \citep{Khandaker}, which is somewhat higher than our new results. 
The calculated thick target yields of the $^{117g,116m,115m,114m,113m,111g}$In radioisotopes are shown in Fig. 21, together with the available literature data. For $^{113m}$In and $^{115m}$In our new data are in excellent agreement with the recent data of \citep{Khandaker}, but for $^{114m}$In our data are lower. No previous data was found for the rest of the radioisotopes presented in Fig. 21. The calculated yields for $^{110m,g109g,108m}$In, $^{115g,111m}$Cd  are shown in Fig. 22. Our data for $^{110g}$In are slightly lower than those of \citep{Khandaker} but in the meta-stable case the data of Khandaker et al. are much higher than ours. These literature data were taken from the IAEA EXFOR database, and if we compare them with the original paper we can recognize that the agreeing data are very different from the graph of the original paper. This confusion might be caused by the fact that in the case of $^{110}$In it is not always clear, which is the ground-state and which is the meta-stable state. For $^{109g}$In there was no literature data found, while for $^{108m}$In the recent data of Khandaker et al. are slightly lower than ours above 20 MeV and slightly higher below this energy. For the radioisotope $^{108g}$In our data agree or in the lower energy region slightly lower than those of \citep{Khandaker}. For $^{115g}$Cd our new data are in acceptable agreement with the recent results of \citep{Khandaker}. Our new calculated yield for $^{111m}$Cd is in good agreement with the data of \citep{Khandaker} above 20 MeV and lower below this energy. 

\begin{figure}
\includegraphics[width=0.5\textwidth]{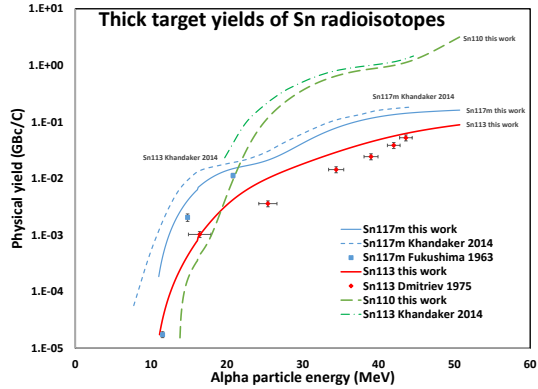}
\caption{Thick target yields of the produced Sn radioisotopes compared with the previous data from the literature}
\label{fig:20}       
\end{figure}

\begin{figure}
\includegraphics[width=0.5\textwidth]{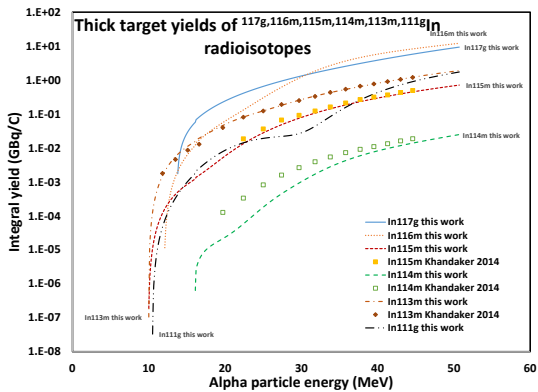}
\caption{Thick target yields of the produced $^{117g,116m,115m,114m,113m,111g}$In radioisotopes compared with the previous data from the literature}
\label{fig:21}       
\end{figure}

\begin{figure}
\includegraphics[width=0.5\textwidth]{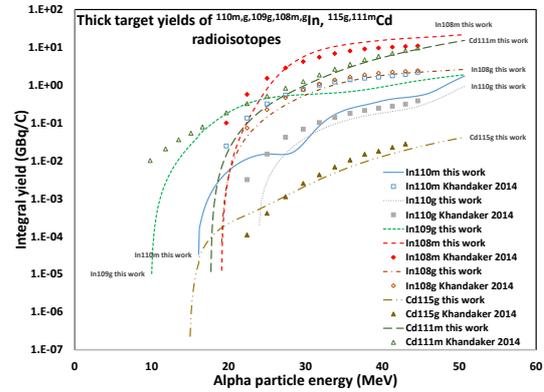}
\caption{Thick target yields of the produced $^{110m,g,109g,108m}$In and $^{115g,111m}$Cd radioisotopes compared with the previous data from the literature}
\label{fig:22}       
\end{figure}

\section{Thin layer activation}
\label{6}
While the cadmium metal is widely used in industry as e.g. alloying element, its application for TLA (Thin Layer Activation) \citep{dit1997, dit2012, dit2006} is also discussed. Thin layer activation can be used for measurement of wear, corrosion and erosion of different surfaces. From the list of the produced isotopes $^{113}$Sn is the most proper choice for thin layer activation because of its convenient half-life (T$_{1/2}$ = 115.09 d), relatively high production cross section (Fig. 4) and the intense gamma-lines (Table 1). Another advantage is that it has local maxima in cross section, which make possible to produce homogeneous activity distribution until a given depth of the surface. Other good candidates are $^{114m}$In and $^{111}$In, which also have still proper nuclear parameters, but have no maximum in the excitation function, which means that only linear activity distribution can be produced. In Fig. 23 the TLA activity distribution curves for $^{113}$Sn are seen, both with 42.1 MeV bombarding energy and 1 hour 2 $\mu$A irradiation parameters. In the first case (15$^o$ irradiation angle) the homogeneity limit (within 1\%) is 10.1 $\mu$m with 80 $\mu$m total penetration depth, while for perpendicular irradiation these values are 39 and 300 $\mu$m, respectively considering 100\% cadmium in the investigated specimen. It demonstrates that the activity distribution can be tuned in wide depth range according to the requirements of the actual wear measurement task. To demonstrate the linear case, the activity distribution of $^{114m}$In is also shown in Fig. 23. The results are calculated with the same irradiation time and beam intensity but with 40 MeV bombarding energy. The linearity limit at the first case (15$^o$ irradiation angle) is 12.2 $\mu$m by 60 $\mu$m total penetration depth and by the perpendicular irradiation these values are 47.2 $\mu$m and 250 $\mu$m, respectively. While by the homogeneous activity distribution the bombarding energy is fixed to an optimum value, by the linear distribution not only the irradiation angle but also the bombarding energy can be varied in order to achieve the desired activity distribution for the particular task.

\begin{figure}
\includegraphics[width=0.5\textwidth]{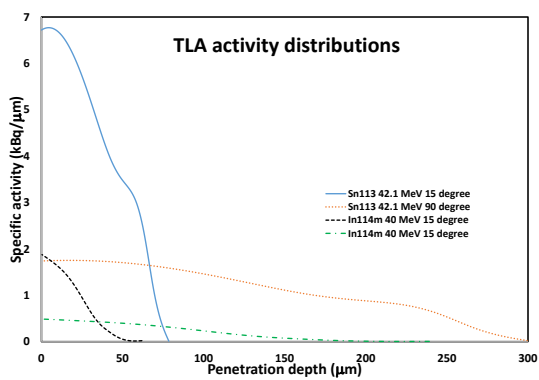}
\caption{TLA activity distributions calculated for $^{113}$Sn and $^{114m}$In with 1 hour 2 $\mu$A irradiations (activity data taken at the End of Bombardment)}
\label{fig:23}       
\end{figure}

\section{Summary and conclusions}
\label{7}

In this study we measured alpha particle induced cross sections for the $^{117m}$Sn, $^{113}$Sn, $^{110}$Sn, $^{117m,g}$In, $^{116m}$In, $^{115m}$In, $^{114m}$In, $^{113m}$In, $^{111}$In, $^{110m,g}$In, $^{109m}$In, $^{108m}$In, $^{115g}$Cd and $^{111m}$Cd radioisotopes by irradiating natural cadmium targets. The new data were compared with the previous data from the literature and with the results of the theoretical nuclear reaction model code calculations with TALYS 1.8 (TENDL-2015) and EMPIRE 3.2 (Malta). In the case of some radioisotopes we extended the energy range of the available data ($^{117g}$In, $^{116m}$In, $^{113m}$In, $^{110m}$In, $^{108m}$In, and $^{111m}$Cd), in other cases tried to dissolve the discrepancy between the existing data sets ($^{117m}$Sn, $^{113}$Sn, $^{111}$In, $^{110m}$In, $^{109g}$In, and $^{115g}$Cd), or we have confirmed the existing data ($^{110}$Sn, $^{117g}$In, $^{116m}$In, $^{115m}$In, $^{114m}$In, $^{113m}$In, $^{110g}$In, $^{108g}$In, and $^{111m}$Cd). These tasks were successfully completed in most cases. The comparison with the model code results shows that their approach by alpha induced reactions, which are more favorable in excitation of high spin isomeric state is not satisfactory. In most cases the data from the TENDL-2015 on-line library \citep{Koning2015b} gave better approximation (e.g. $^{113}$Sn, $^{110}$Sn, etc.), but there were cases where EMPIRE 3.2 delivered better approximation (e. g. $^{108g}$In), but in one case both failed completely ($^{111m}$Cd). For comparison the thick target physical yields were calculated from the experimental excitation functions and compared with the available literature data. The data in Figs. 20-22 show an acceptable agreement in most cases. A serious disagreement was seen only in the case on $^{110m}$In, where also a possible explanation is given. The applicability of some produced radioisotopes for wear measurement by using thin layer activation is also demonstrated.

\section{Acknowledgements}
\label{}
This work was performed at the RI Beam Factory operated by the RIKEN Nishina Center and CNS, University of Tokyo. This work was carried out in the frame of the standing HAS-JSPS (Hungary Japan) bilateral exchange agreement. The authors acknowledge the support of the respective institutions in providing technical support and use of experimental facilities. (Contract No.: NKM-89/2014).
 



\bibliographystyle{elsarticle-harv}
\bibliography{Cdnat}







\end{document}